\documentclass{article}
\usepackage{spconf,amsmath,graphicx,hyperref, epsfig, amsfonts}
\usepackage{cite}

\title{Pre-training Tensor-Train Networks Facilitates Machine Learning with Variational Quantum Circuits}
\name{Jun Qi$^1$, Chao-Han Huck Yang$^2$, Pin-Yu Chen$^3$, Min-Hsiu Hsieh$^4$}
\address{1. Georgia Institute of Technology, Atlanta, GA, USA \\
	      2. NVIDIA Research, CA, USA \\
	      3. IBM Research, NY, USA  \\
	      4. Hon Hai (Foxconn) Quantum Computing Research Center, Taipei, Taiwan   \\}
\begin{document}
%
\maketitle
\begin{abstract}
Data encoding remains a fundamental bottleneck in quantum machine learning, where amplitude encoding of high-dimensional classical vectors into quantum states incurs exponential cost. In this work, we propose a pre-trained tensor-train (TT) encoding network (Pre-TT-Encoder) that significantly reduces the computational complexity of amplitude encoding while preserving essential data structure. The Pre-TT-Encoder exploits low-rank TT decompositions learned from classical data, enabling polynomial-time state preparation in the number of qubits and TT-ranks. We provide a theoretical analysis of the encoding complexity and establish fidelity bounds that quantify the trade-off between TT-rank and approximation error. Empirical evaluations on classical (MNIST) and quantum-native (semiconductor quantum dot) datasets demonstrate that our approach achieves substantial gains in encoding efficiency over direct amplitude encoding and PCA-based dimensionality reduction, while maintaining competitive performance in downstream variational quantum circuit classification tasks. The proposed method highlights the role of tensor networks as scalable intermediaries between classical data and quantum processors. 
\end{abstract}
\section{Introduction}
\label{sec:intro}

Quantum machine learning (QML)~\cite{biamonte2017quantum, caro2022generalization, power_data} has emerged as a promising paradigm that leverages the computational power of quantum processors to enhance classical learning tasks. Among different architectures, variational quantum circuits (VQCs)~\cite{qi2023theoretical, qi2025tensorhyper, qi2023qtn} play a central role due to their suitability for noisy intermediate-scale quantum (NISQ) devices~\cite{preskill2018quantum, bharti2022noisy}. A critical component in VQC-based models is the encoding of classical data into quantum states~\cite{schuld2021effect, schuld2018supervised}, which serves as the interface between high-dimensional classical features and quantum computing. 

Despite its importance, data encoding remains a significant bottleneck in QML. In particular, amplitude encoding, which maps a classical vector of dimension $2^{U}$ into the amplitudes of a $U$-qubit quantum state, offers optimal qubit efficiency but incurs exponential classical preprocessing cost~\cite{cerezo2022challenges, huang2021power}. Existing dimensionality reduction techniques, such as principal component analysis (PCA)~\cite{abdi2010principal}, alleviate this burden but often lose task-relevant structure and do not scale well to high-dimensional or structured data~\cite{qi2023theoretical}. Consequently, efficient and faithful encoding methods are required to bridge the gap between classical datasets and quantum hardware. 

To address this challenge, we propose a pre-trained tensor-train encoding network (Pre-TT-Encoder) that exploits the expressive yet efficient tensor-train (TT) decomposition. The TT-format factorizes high-order tensors into a chain of low-rank cores, enabling polynomial storage and computational complexity with respect to the tensor rank rather than the ambient data dimension~\cite{oseledets2011tensor, yang2017tensor, qi2023exploiting}. By pre-training the TT decomposition on classical datasets, our approach captures domain-specific structure while yielding a compressed representation that can be directly mapped to quantum states through amplitude encoding. This significantly lowers the computational cost of state preparation, making amplitude encoding more practical on NISQ devices.

Beyond practical efficiency, the Pre-TT-Encoder has strong theoretical underpinnings. We analyze its encoding complexity, showing that TT decomposition reduces the cost of amplitude encoding from $\mathcal{O}(2^{U})$ to $\mathcal{O}(Ur^2)$, where $r$ denotes the TT-rank. We further provide fidelity guarantees, demonstrating that the approximation error~\cite{bach2024learning, mohri2018foundations, qi2020analyzing} introduced by low-rank TT compression remains bounded and controllable. 

We validate our method through experiments on both classical datasets (MNIST digit classification~\cite{deng2012mnist}) and quantum-native datasets (semiconductor quantum dot classification~\cite{gualtieri2025qdsim}). Results show that the Pre-TT-Encoder achieves substantial improvements in encoding efficiency compared to amplitude encoding and PCA-based preprocessing, while preserving or improving classification accuracy in VQC-based models. Our contributions are summarized as follows: 
\begin{enumerate}
\item We identify amplitude encoding as a computational bottleneck in QML and propose the Pre-TT-Encoder, a tensor-train-based framework for efficient quantum state preparation. 
\item We provide a theoretical analysis of complexity and fidelity trade-offs in TT-based amplitude encoding. 
\item We empirically demonstrate the effectiveness of our approach across classical and quantum-native datasets, highlighting both efficiency gains and improvements in downstream learning performance. 
\end{enumerate}

\section{Background and Related Work}

\subsection{Variational Quantum Circuits}

\begin{figure}[htp]
\centerline{\epsfig{figure=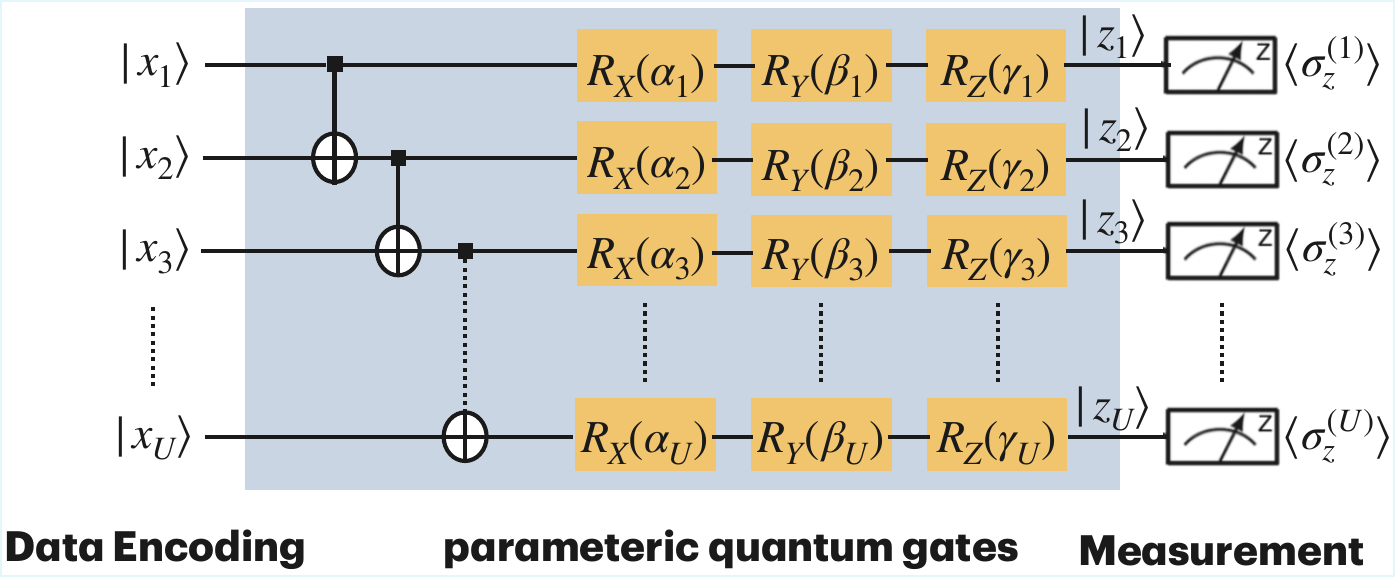, width=66mm}}
\caption{{\it The architecture of a variational quantum circuit}. The circuit consists of three main components: (i) data encoding, where classical data are mapped to quantum states $\vert x_{1} \rangle$, $\vert x_{2} \rangle$, ..., $\vert x_{U} \rangle$; (ii) parametric quantum gates, including rotation gates $R_{X}(\alpha_u)$, $R_{Y}(\beta_{u})$, $R_{Z}(\gamma_{u})$, which are trained to optimize the model; and (iii) measurement, where observables $\langle \sigma_{z}^{(1)}\rangle$, $\langle \sigma_{z}^{(2)}\rangle$, ..., $\langle \sigma_{z}^{(U)} \rangle$ are evaluated to produce classical outputs for learning tasks.} 
\label{fig:qnn}
\end{figure}

VQCs are parameterized quantum circuits trained with classical optimizers to minimize task-specific cost functions. VQCs have been applied to classification, regression, and generative modeling, serving as the foundation for near-term QML on NISQ devices. A VQC typically consists of three stages: (i) data encoding, where classical vectors are mapped into quantum states; (ii) a sequence of parameteric quantum gates with trainable parameters~\cite{benedetti2019parameterized}; and (iii) measurement and classical post-processing. Among them, the data encoding stage is often the computational bottleneck, especially for high-dimensional classical datasets. 

\subsection{Amplitude Encoding}

Amplitude encoding encodes a normalized classical vector $\textbf{x} \in \mathbb{R}^{2^{U}}$ into a $U$-qubit quantum state: 
\begin{equation}
\vert \textbf{x} \rangle = \sum\limits_{u=0}^{2^{U} - 1} x_{u} \vert u \rangle, 
\end{equation}
which is maximally efficient in terms of qubit usage. However, preparing such a state typically requires $\mathcal{O}(2^{U})$ operations, making it exponentially costly for large input dimensions. Alternative encodings, such as angle encoding or data re-uploading, avoid this exponential but require more qubits or deeper circuits, thereby introducing noise and trainability issues. Efficient and structure-preserving amplitude encoding remains a central challenge for scalable QML. 

\subsection{Tensor-Train Networks}
TT networks, also known as matrix product states (MPS)~\cite{cirac2021matrix, perez2006matrix, orus2014practical}, provide a compact factorization of high-order tensors into a sequence of low-rank cores. A vector of dimension $2^{U}$ can be reshaped into a high-order tensor and represented in TT form, with complexity reduced to $\mathcal{O}(Ur^2)$, where $r$ is the TT-rank. TT networks have been widely applied to compress deep learning models, simulate quantum many-body systems~\cite{ran2020tensor}, and improve learning efficiency in hybrid quantum-classical models. Their ability to capture global correlations with polynomial complexity makes them natural candidates for efficient data encoding in QML. 

\subsection{Related Work in QML Encoding}

Several approaches have been proposed to mitigate encoding overhead in QML. PCA-based preprocessing reduces data dimensionality before encoding, but it often discards information relevant to downstream tasks. Tensor networks have been explored for quantum state simulation and feature compression, though typically without explicit pretraining to improve encoding efficiency. More recently, data re-uploading schemes and quantum kernel methods~\cite{jerbi2023quantum} have been studied as an alternative to direct amplitude encoding. However, these approaches either require deeper circuits or incur high computational costs from classical kernel evaluation~\cite{bousquet2002complexity}. 

\subsection{Distinction of This Work}
Our approach departs from prior studies by introducing a pre-trained tensor-train encoder (Pre-TT-Encoder) that directly targets the computational complexity of amplitude encoding. Unlike PCA, it preserves structured correlations in data through low-rank factorization. Unlike generic tensor networks, it is trained on classical datasets to capture domain-specific features before quantum state preparation. 

\section{Pre-TT-Encoder Framework}

\subsection{Overview}
The proposed Pre-TT-Encoder mitigates the exponential cost of amplitude encoding by leveraging a pre-trained TT network as a structured dimensionality-reduction module. The framework consists of two main stages: 
\begin{enumerate}
\item \textbf{Pre-training Stage}: A TT decomposition is trained on classical data to extract low-rank structure. 
\item \textbf{Quantum State Preparation Stage}: The TT representation is mapped into qubit amplitudes for efficient state preparation. 
\end{enumerate}

This pipeline transforms the naive $\mathcal{O}(2^{U})$ cost of encoding into $\mathcal{O}(Ur^{2})$, where $U$ is the number of qubits and $r$ is the TT-rank.

\subsection{Pre-training Stage}
Given a dataset $\mathcal{D} = \{\textbf{x}^{(m)} \in \mathbb{R}^{D}, \textbf{y}^{(m)}\}$, we first reshape each input vector $\textbf{x}^{(m)}$ into a tensor as:
\begin{equation}
\mathcal{X} \in \mathbb{R}^{I_{1} \times I_{2} \times \cdot\cdot\cdot \times I_{K}}, \hspace{2.5mm} \prod\limits_{k=1}^{K} I_{k} = 2^{U}, 
\end{equation}
where $K$ is the chosen TT order. 

The TT decomposition factorizes $\mathcal{X}$ into a product of TT-cores:
\begin{equation}
\mathcal{X}(i_1, i_2, ..., i_K) = \mathcal{G}_{1}(i_1) \mathcal{G}_{2}(i_2)\cdot\cdot\cdot\mathcal{G}_{K}(i_{K}), 
\end{equation}
where each TT-core $\mathcal{G}_{k}(i_{k}) \in \mathbb{R}^{r_{k} \times r_{k+1}}$, and the TT-ranks $\{r_{k}\}$ control compression and are bounded by $r = \max_{k} r_{k}$.  

The TT is pre-trained either supervisedly (e.g., by minimizing reconstruction error) or unsupervisedly (e.g., by predicting labels). Once trained, the TT parameters are frozen, ensuring stable, low-rank feature representations.

Moreover, the cost of contracting TT-cores scales as: 
\begin{equation}
\mathcal{O}\left( \sum\limits_{k=1}^{K} I_{k}r_{k}r_{k+1} \right) \le \mathcal{O}\left(K r^{2} \max_{k} I_{k} \right),
\end{equation}
which is polynomial in $K$ and $r$. Choosing $K = U$ (splitting into single-qubit dimensions $I_{k}=2$) yields $\mathcal{O}(Ur^{2})$.

\subsection{Quantum State Preparation Stage}
The output of the pre-trained TT operator $\psi_{\rm tt}(\cdot)$ is a compressed vector $\hat{\textbf{x}} \in \mathbb{R}^{2^{U}}$, where $U$ is the number of qubits available for amplitude encoding. This compressed representation is normalized to form a quantum state: 
\begin{equation}
\vert \psi_{\rm tt}(\textbf{x}) \rangle = \frac{1}{\lVert \hat{\textbf{x}} \rVert_{2}} \sum\limits_{u=0}^{2^{U} - 1} \hat{x}_{u} \vert u \rangle. 
\end{equation}

Crucially, the TT decomposition allows efficient computation of the amplitude $\hat{x}_{u}$ by contracting a sequence of small TT-cores. The complexity reduces from $\mathcal{O}(2^{U})$ to $\mathcal{O}(Ur^{2})$, enabling scalable encoding for moderate qubit counts.

\subsection{Fidelity Analysis}
When compressing high-dimensional data, TT decomposition introduces approximation error. Let the original vector be $\textbf{x}$ and the TT-compressed approximation be $\hat{\textbf{x}}$. The truncation error is defined as: 
\begin{equation}
\lVert \textbf{x} - \hat{\textbf{x}} \rVert_{2} \le \epsilon. 
\end{equation}

The corresponding encoded states are 
\begin{equation}
\vert \textbf{x} \rangle = \frac{\textbf{x}}{\lVert \textbf{x} \rVert_{2}}, \hspace{2mm} \vert \psi_{\rm tt}(\textbf{x})\rangle  = \frac{\hat{\textbf{x}}}{\lVert \hat{\textbf{x}} \rVert_{2}}. 
\end{equation}

The fidelity between the exact and approximate states $\mathcal{F}(\textbf{x}, \hat{\textbf{x}})$ is given by
\begin{equation}
\mathcal{F}(\textbf{x}, \hat{\textbf{x}}) = \vert \langle \textbf{x} \vert \psi_{\rm tt}(\textbf{x}) \rangle \vert^{2}. 
\end{equation} 

Using the perturbation bound, we obtain:
\begin{equation}
\mathcal{F}(\textbf{x}, \hat{\textbf{x}}) \ge 1 - \mathcal{O}\left( \frac{\epsilon^{2}}{\lVert \textbf{x} \rVert_{2}^{2}} \right). 
\end{equation}

This result indicates that when the TT-rank is sufficiently large, the approximation error $\epsilon$ becomes negligible, leading to fidelities approaching unity. Conversely, reducing the TT-rank enhances computational efficiency at the expense of fidelity. Thus, the Pre-TT-Encoder provides a flexible mechanism to balance encoding complexity and state fidelity, allowing practitioners to adapt the trade-off to specific application requirements.

\section{Experiments}

This section evaluates the proposed Pre-TT-Encoder against two baselines, including Amplitude Encoding (AE) and PCA+VQC, on (i) a standard handwritten digit classification and (ii) a Quantum Dot classification task. We report accuracy, cross-entropy loss, and end-to-end wall-clock time. 

In particular, all models use the same hardware-efficient VQC ansatz, consisting of $4$ layers of parameterized single-qubit rotations $(R_X, R_Y, R_Z)$ followed by a CNOT ring entangling all qubits. Training uses cross-entropy loss with the Adam optimizer~\cite{adam:2014} at a learning rate of $3\times 10^{-3}$ and a batch size of $64$. The pre-trained TT network is obtained from our previous work in~\cite{qi2023exploiting}.

\subsection{Handwritten Digit Classification}
We adopt a standard binary subset of the MNIST dataset by selecting digits $2$ and $5$, which are difficult to discriminate because of their structural similarity. Each grayscale image ($28\times 28$) is flattened to a $784$-dimensional vector. Since amplitude encoding requires the input dimension to be a power of two, vectors are zero-padded to $2^{10} = 1024$, corresponding to $10$ qubits. The train/test split follows the original MNIST partition, with a small subset of the training set used for faster benchmarking.

\begin{table}[h]
\centering
\caption{Comparison of encoding methods on the subset of the MNIST dataset (2 vs 5).}
\begin{tabular}{lccc}
\hline
Method & Test loss & Test acc. $(\%)$ & Time (s) \\
\hline
AE & 0.1992 & 0.920998 & 142.714 \\
\hline
PCA+VQC & 0.4269 & 0.796778 & 149.502 \\
\hline
\textbf{Pre-TT-Encoder} & \textbf{0.1514} & \textbf{0.988129} & \textbf{132.633} \\
\hline
\end{tabular}
\end{table}

On the MNIST $2$ vs. $5$ classification task, the empirical results demonstrate a clear advantage of the proposed Pre-TT-Encoder over both amplitude encoding and PCA-based encoding. The Pre-TT-Encoder achieved a best test accuracy of $98.8\%$, which is markedly higher than the $92.1\%$ reached by amplitude encoding and the $79.7\%$ obtained with PCA+VQC. In addition, the corresponding cross-entropy loss for Pre-TT-Encoder was the lowest at $0.1514$, compared to $0.1992$ for amplitude encoding and $0.4269$ for PCA+VQC, highlighting the improved fidelity of the TT-based representation. 

More importantly, this performance gain did not come at the expense of efficiency. The total training time was $132.6$ seconds, outperforming amplitude encoding (142.7 seconds) and PCA+VQC (149.5 seconds). These results indicate that the Pre-TT-Encoder simultaneously enhances accuracy, reduces information loss, and lowers runtime, validating its effectiveness as a scalable encoding strategy for variational quantum classifiers.

\subsection{Quantum Dot Classification}

We next evaluated the three encoding strategies on the quantum dot classification dataset, which comprises $50 \times 50$ pixel images representing quantum dot charge stability diagrams~\cite{czischek2021miniaturizing, tang2015storage, brennan2011atomic}, including $2,000$ diagrams generated under realistic experimental noise conditions. We randomly partition the noiseless diagrams into $1,800$ training and $200$ test samples to evaluate the models' performance. Notably, this task differs from MNIST in that the input distributions are smooth and highly correlated across neighboring wavelengths, thereby exhibiting strong low-rank structure. Such a structure is expected to benefit from tensor-based encodings. 

\begin{table}[h]
\centering
\caption{Comparison of encoding methods on the quantum dot classification task.}
\begin{tabular}{lccc}
\hline
Method & Test loss & Test acc. $(\%)$ & Time (s) \\
\hline
AE               & 0.731871 & 0.4825 & 260.540 \\
\hline
PCA+VQC          & 0.492000 & 0.7775 & 274.965 \\
\hline
\textbf{Pre-TT-Encoder} & \textbf{0.320339} & \textbf{0.8725} & \textbf{247.174} \\
\hline
\label{tab:qd_results}
\end{tabular}
\end{table}

The empirical results, reported in Table~\ref{tab:qd_results}, confirm this intuition. Amplitude Encoding struggled, achieving only $48.3\%$ accuracy with a relatively high test loss of $0.732$, suggesting that directly embedding the spectrum into amplitudes without structure-aware compression leads to poor generalization. PCA+VQC improved performance to $77.8\%$ accuracy with a lower test loss of $0.492$, highlighting the benefits of linear dimensionality reduction. However, the linearity of PCA limits its ability to capture multi-way correlations inherent in spectral data. 

In contrast, the proposed Pre-TT-Encoder achieved substantially higher accuracy of $87.3\%$ with the lowest test loss of $0.320$, while also reducing runtime to $247.2$ seconds, compared with amplitude encoding ($260.5$ seconds) and PCA+VQC ($275.0$ seconds). These gains reflect TT truncation's ability to retain the most informative correlations in the spectral input while discarding redundancy and noise. The result is a compressed amplitude state that is both more faithful to the data's intrinsic structure and more efficient to prepare for the VQC. 

\subsection{Discussions}
Based on the two experimental results above, the Pre-TT-Encoder provides a superior trade-off among fidelity, generalization, and efficiency. By leveraging low-rank structure in scientific spectra, it outperforms both conventional amplitude encoding and classical linear compression, suggesting its promise as a practical encoding scheme for real-world quantum machine learning applications.

\section{Conclusions}

In this work, we addressed the longstanding challenge of efficient amplitude encoding in quantum machine learning. We proposed the Pre-TT-Encoder, a tensor-train–based framework that leverages pre-trained low-rank decompositions to reduce the exponential cost of state preparation. Our theoretical analysis established that the computational complexity can be reduced from $\mathcal{O}(2^{U})$ to $\mathcal{O}(Ur^{2})$, with fidelity guarantees that quantify the trade-off between TT-rank and approximation accuracy. 

Empirical evaluations on both classical (MNIST) and quantum-native (quantum dot spectra) datasets demonstrated the advantages of our method. The Pre-TT-Encoder not only achieved higher accuracy than amplitude encoding and PCA+VQC, but also reduced runtime, highlighting its effectiveness as a structure-aware, scalable encoder for VQCs. 

In summary, our theoretical and empirical results suggest that TT–guided preprocessing offers a promising approach to bridge classical data and near-term quantum devices. 

\section{Knowledgement}
This work is jointly funded by the Hong Kong Research Impact Fund (R6010-23) and the Hong Kong Research Grants Council Fund (12203124). 

\vfill\pagebreak

\small
\bibliographystyle{IEEEbib}
\bibliography{sn-bibliography}

\end{document}